\documentclass[aps,showpacs,amsmath,amssymb,pre,superscriptaddress,twocolumn]{revtex4}
\usepackage{graphicx}
\usepackage{epsfig}
\usepackage{ulem}
\usepackage{amsmath}
\usepackage{amssymb}
\usepackage{color}

\begin{document}

\title{Reducing influence of antiferromagnetic interactions on ferromagnetic properties of p-(Cd,Mn)Te quantum wells}

\author{C. Simserides}
\email{csimseri@phys.uoa.gr}
\affiliation{Institute for Advanced Materials, Physicochemical Processes, Nanotechnology and Microsystems, NCSR Demokritos, GR-15310 Athens, Greece}
\affiliation{Present address: Physics Department, University of Athens, GR-15784 Athens, Greece} 
\author{A. Lipi\'{n}ska}
\affiliation{Institute of Physics, Polish Academy of Sciences, PL-02 668 Warszawa, Poland}
\author{K.N. Trohidou}
\affiliation{Institute for Advanced Materials, Physicochemical Processes, Nanotechnology and Microsystems, NCSR Demokritos, GR-15310 Athens, Greece}
\author{T. Dietl}
\affiliation{Institute of Physics, Polish Academy of Sciences, PL-02 668 Warszawa, Poland}
\affiliation{Institute of Theoretical Physics, Faculty of Physics, University of Warsaw, PL-00 681 Warszawa, Poland}
\affiliation{WPI-Advanced Institute for Materials Research, Tohoku University, Sendai 980-8577, Japan}

\date{\today}

\begin{abstract}
In order to explain the absence of hysteresis in ferromagnetic p-type (Cd,Mn)Te quantum wells (QWs),
spin dynamics was previously investigated by Monte Carlo simulations combining the Metropolis algorithm with
the determination of hole eigenfunctions at each Monte Carlo sweep.
Short-range antiferromagnetic superexchange interactions between Mn spins
--which compete with the hole-mediated long-range ferromagnetic coupling--
were found to accelerate magnetization dynamics if the the layer containing Mn spins is wider than the vertical range of the hole wave function.
Employing this approach it is shown here that appreciate magnitudes of remanence and coercivity can be obtained if
Mn ions are introduced to the quantum well in a delta-like fashion.
\end{abstract}


\maketitle

\section{Introduction}
\label{sec:introduction}
Ferromagnetism in (Cd,Mn)Te/(Cd,Mg,Zn)Te modulation-doped $p$-type quantum wells (QWs) was revealed
by the observation of spontaneous splitting of the fundamental photoluminescence line below a certain temperature $T_{\mathrm{C}}$ \cite{Haury:1997,Kossacki:2000,Boukari:2002}.
The magnitude of $T_{\mathrm{C}}$ as a function of the Mn concentration $x$ and the hole areal density $p$
can be quantitatively explained if in addition to carrier-mediated ferromagnetic (FM) couplings,
the presence of competing short-range antiferromagnetic (AFM) superexchange interactions as well as
a Stoner enhancement factor are taken into account \cite{Haury:1997,Boukari:2002,Dietl:1997,Dietl:1999,Kechrakos:2005}.

A surprising finding of optical and magnetic measurements is the absence of hysteresis loops and,
hence, of macroscopic spontaneous magnetization below $T_{\mathrm{C}}$ \cite{Kossacki:2000,Kossacki:2002}.
In order to shed some light on this issue, earlier time-resolved magneto-optical measurements \cite{Kossacki:2002} and
Monte Carlo (MC) simulations~\cite{Kechrakos:2005}, have been extended by the recent more comprehensive studies \cite{LSTGKMD:2009}.
According to the experimental findings \cite{LSTGKMD:2009}, above $T_{\mathrm{C}}$ the magnetization relaxation time is shorter than 20~ns.
Below $T_{\mathrm{C}}$ a critical slowing-down is observed.
Nevertheless, magnetization correlation persists only up to a few $\mu$s,
the time scale consistent with the absence of spontaneous magnetization in the static measurements.
We found a similar relative elongation of the relaxation time below $T_{\mathrm{C}}$
by MC simulations based on the Metropolis algorithm and the determination of one-particle hole eigenstates at each MC sweep \cite{LSTGKMD:2009}.
We demonstrated also that short range AFM interactions
are crucial in accelerating magnetization dynamics and
that the influence of AFM interactions can be diminished
if the thickness of the layer containing Mn spins is smaller than the vertical range of the hole wave function.
This finding confirmed theoretical considerations of Boudinet and Bastard~\cite{Boudinet:1993}
that magnetization relaxation of bound magnetic polarons in $p$-type CdTe/(Cd,Mn)Te QWs occurs
due to the AFM coupling to Mn spins located outside the relevant Bohr radius.
Moreover, we studied how a spin-independent part of the potential introduced by the Mn impurities affects $T_{\mathrm{C}}$ \cite{LSTGKMD:2009}.
Our simulations showed that alloy disorder tends to reduce $T_{\mathrm{C}}$.
The effect is particularly dramatic for the attractive alloy potential which,
if sufficiently large, leads to a strong hole localization.
This result substantiates the notion that delocalized or weakly localized carriers are necessary
to generate a sizable FM coupling between diluted localized spins.

In this paper, we investigate the form of hysteresis loops in the case when Mn ions
are introduced to the quantum well in a delta-like fashion.
Our results suggest that large magnitudes of remanence and coercivity can be expected in such samples
without any significant lowering of $T_{\mathrm{C}}$.

\section{Theory}
\label{sec:theory}
As discussed previously \cite{LSTGKMD:2009},
our approach encompasses automatically the description of carrier-mediated exchange interactions within
either $p$-$d$ Zener model or  Ruderman-Kittel-Kasuya-Yosida theory.
The carrier total carrier energy is evaluated at given Mn spin configuration neglecting hole correlations,
which underestimates the ferromagnetic coupling~\cite{Dietl:1997}.
Thus, our computed $T_{\mathrm{C}}$ is systematically lower than the experimental one.
We assume that the in-plane and the perpendicular hole motions can be factorized, so that
the hole envelope function assumes the form,
$\Psi_{\sigma}(\textbf{R}) = \psi_{\sigma}(\textbf{r})\varphi(z)$.
$\textbf{R} = (\textbf{r},z)$, where
$\textbf{r} = (x,y)$ and $z$ are the in-plane and perpendicular hole coordinates, respectively.
For the ground state heavy-hole subband wave function $\varphi(z)$
that corresponding to an infinite QW is used.
Spin-orbit interaction fixes the hole spin along the growth axis.

Without an external magnetic field, the Hamiltonian reads
\begin{equation}\label{htotal}
\mathcal{H} = \frac{p^2}{2m^*} + \mathcal{H}_{pd} + \mathcal{H}_{dd}.
\end{equation}
$m^{*} = 0.25m_{0}$ \cite{Fishman:1995}.
We take the $p$-$d$ interaction in the form
\begin{equation}\label{hpd}
\mathcal{H}_{pd} = \sum_{i} \frac{1}{3} \beta j_{z} S_{zi}
\delta(\textbf{r}- \textbf{r}_{i}) \left|\varphi(z_{i})\right|^{2},
\end{equation}
where the hole spin $j = 3/2$, $j_{z} = \pm $ 3/2.
The Mn ions are randomly distributed over one of the two fcc lattices
which make up the zinc blende crystal structure.
Their positions are denoted by $\textbf{r}_{i}$.
We treat Mn spins $\textbf{S}_{i}$ classically, with $S = 5/2$.
This is qualitatively valid for the large spin in question,
but quantitatively --according to mean-field theory--
it underestimates $T_{\mathrm{C}}$ by $(S+1)/S$.
The $p$-$d$ exchange integral $\beta =  -5.96 \times 10^{-2}$~eV~nm$^3$, i.e.,
the exchange energy $\beta N_0 = -0.88$~eV~\cite{Gaj:1979}, where
$N_0$ is the cation concentration.
In the previous work \cite{LSTGKMD:2009}, to include a {\it spin-independent} alloy disorder introduced by the Mn ions,
we added to Eq.~\ref{htotal}  a term
\begin{equation}\label{hAD}
\mathcal{H}_{\mathrm{AD}} = \delta V \sum_{i} \delta(\textbf{r}- \textbf{r}_{i})\left|\varphi(z_{i})\right|^{2},
\end{equation}
where $\delta V = 3.93 \times 10^{-2}$~eV~nm$^3$,
as the valence band offset $W N_0 = -\delta V N_0 = -0.58$~eV in (Cd,Mn)Te~\cite{Gaj:1994,Wojtowicz:1996}.
The effect of the Mn impurities on the carrier wave function is governed by ~\cite{Dietl:2008}
\begin{equation}\label{uperuc}
U/U_c = 6m^*[W - (S + 1)\beta/2]/(\pi^3\hbar^2b),
\end{equation}
where $U$ is the total Mn potential and
$U_c <0$ is its critical value at which a bound state starts to form and
$b$ is the potential radius.
The Mn polarization $\langle S_z \rangle$ is typically below 10\%.
Thus, for $S+1 \rightarrow \langle S_z\rangle$ the total Mn potential
is expected to be effectively repulsive for holes in (Cd,Mn)Te, $U/U_c <0$.
By using $\delta$-like potentials,
which corresponds to $b\rightarrow 0$, we actually overestimated $U/U_c$ and
determined an upper limit of the effect of the alloy potential upon $T_{\mathrm{C}}$ \cite{LSTGKMD:2009}.
The short-range AFM interaction between Mn spins is given by
\begin{equation}\label{hdd}
\mathcal{H}_{dd} = -2k_{B} \sum_{ij}  J_{ij} \textbf{S}_{i} \cdot \textbf{S}_{j},
\end{equation}
\noindent
where $J_{ij} = -6.3, -1.9, -0.4$~K for the nearest, next nearest, and next next nearest neighbors, respectively~\cite{Shapira:2002}.
When a magnetic field is applied,
we add to Eq.~\ref{htotal} the Zeeman coupling
between the hole spin and the magnetic field
as well as
between the Mn spin and the magnetic field.

We adopt a simulation box with dimensions $L \times L \times L_{\mathrm{W}}$.
$L = 350a_o$ ($x$- and $y$-axis) and $L_{W} = 8a_o$ ($z$-axis). The  fcc lattice constant $a_o = 0.647$~nm.
Hence, the QW width is $L_{\mathrm{W}} = 5.2$~nm and the area $L^2 = $(226~nm)$^2$.
A typical QW width of experimental samples is somewhat greater, $L_{\mathrm{W}}^{exp} = 8$~nm.
The adoption of a smaller value of $L_{\mathrm{W}}$ allows us to treat systems with a larger area and,
at the same time, since within the mean-field theory~\cite{Dietl:1997} $T_{\mathrm{C}} \propto 1/L_{\mathrm{W}}$,
results in a partial compensation of a systematic error resulting from treating the spins classically,
$L_{\mathrm{W}}^{exp}S/[L_{\mathrm{W}}(S+1)] = 1.1$.
To determine hole eigenstates for a given configuration of Mn spins
we assume periodic boundary conditions in the QW atomic layers and diagonalize $\mathcal{H}$
in a plane-wave basis with 2D wave vectors truncated~\cite{Schliemann:2001}
at a radius $k_c = 8.10 (2\pi/L)$.
This part of the 2D $k$-space is sufficiently large to ensure convergence for all $p$ considered here.
The energy of the holes is determined by summing up the lowest eigenvalues corresponding to a given number of holes $N_\mathrm{h}$.
To keep $k$-space shells filled, we take $N_{\mathrm{h}} = $ 1, 5, 9, 13, 21, 25, 29, 37, 45, 49, and 57,
i.e., for the chosen size of the simulation box, $p \leq 1.11 \times 10^{11}$~cm$^{-2}$.
We employ the Metropolis algorithm in which the hole eigenfunctions and eigenvalues are updated at each MC sweep
(in one MC sweep all Mn spins are rotated).
This procedure if applied after each single spin rotation would be computationally excessively time consuming and,
therefore, we followed the idea of the so-called
perturbative Monte Carlo method~\cite{Duane:1987,Troung:1996,Kennett:2002}.
We typically keep 2000 initial MC sweeps to thermalize the system,
followed by $10^{4}$-$10^{5}$ MC sweeps for further analysis.
Hence, we need a (pseudo)random number generator with a {\it long} period \cite{LSTGKMD:2009}.

We calculate the temperature ($T$) dependence of the spin projections
and the spin susceptibilities \cite{Newman:1999} both for holes and magnetic ions.
For spins we denote $\sigma = s$ (holes) and $\sigma = S$ (Mn ions).
Similarly, $N = N_\mathrm{h} \;\; (N_\mathrm{Mn})$ for the number of holes (Mn ions).
At each MC sweep, the spin projections per hole or per Mn ion are given by:
$\overline{\sigma_j} = (\sum_{i=1}^N \sigma_{ij})/N, \hspace{0.2cm} j=x,y,z.$
We denote statistical averages by $\langle ... \rangle$:
$\langle \sigma_j \rangle   = (\sum_{n=1}^{n_t}\overline{\sigma_j})/n_t$,
$\langle |\sigma_j| \rangle = (\sum_{n=1}^{n_t}|\overline{\sigma_j}|)/n_t,$
where $n$ ($n_t$) denotes successive (the total number of) MC sweeps used for the statistical average.

\section{Results and discussion}
\label{sec:results}
In order to find out how AFM interactions affect magnetization dynamics
we calculated \cite{LSTGKMD:2009} the magnetization autocorrelation function \cite{Newman:1999}
with and without AFM interactions at $T = 0.7 T_{\mathrm{C}}$.
Our results indicated that short-range AFM interactions strongly accelerate magnetization fluctuations and
account for the absence of hysteresis in static measurements.
We repeated \cite{LSTGKMD:2009} calculations with Mn ions occupying
only the central part of the QW, the thickness of the Mn layer being half of the QW width $L_{\mathrm{W}}$.
We found that in such a case magnetization dynamics resembles the case when AFM interactions are switched off.
We concluded, therefore, that Mn spins close to the QW edges account for fast relaxation of magnetization.

\begin{table}[h!]
\caption{Remanent spin of Mn ions and holes
($| \langle S_z \rangle |$ and $| \langle s_z \rangle |$, respectively, for $B = $ 0),
obtained by Monte Carlo simulations at 0.2 K
for  $p$-type quantum wells (QWs) with various Mn content $x$ and assuming Mn cations to be randomly distributed
either uniformly over the QW or in a narrow region in the center of the QW.
The number of holes is $N_h = 13$, which corresponds to areal hole density $p =  2.54 \times 10^{10}$ cm$^{-2}$.}
\vspace{0.2cm}
\centerline{
\begin{tabular}{|l|l|r|l|} \hline
Remanence    & Remanence    &   x \%  &  Mn-doping shape  \\
of Mn ions   & of holes     &         &                   \\ \hline  \hline
0.098        & 0.5          &   4     &  central delta    \\ \hline
0.044        & 0.5          &   4     &  uniform          \\ \hline
0.061        & 0.5          &   8     &  central delta    \\ \hline
0.019        & 0.5          &   8     &  uniform          \\ \hline
0.023        & 0.42         &  16     &  central delta    \\ \hline
\end{tabular}}
\label{table:non-delta_delta}
\end{table}

\begin{figure}[t!]
\begin{center}
\includegraphics[scale=0.6]{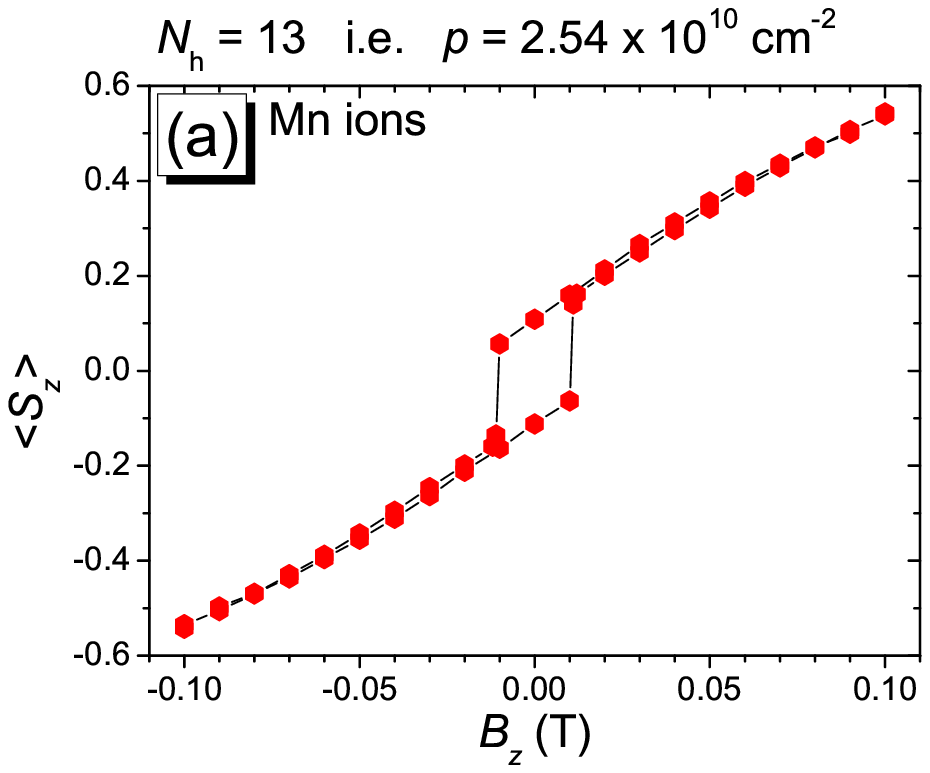}
\vspace{-0.75cm}
\includegraphics[scale=0.6]{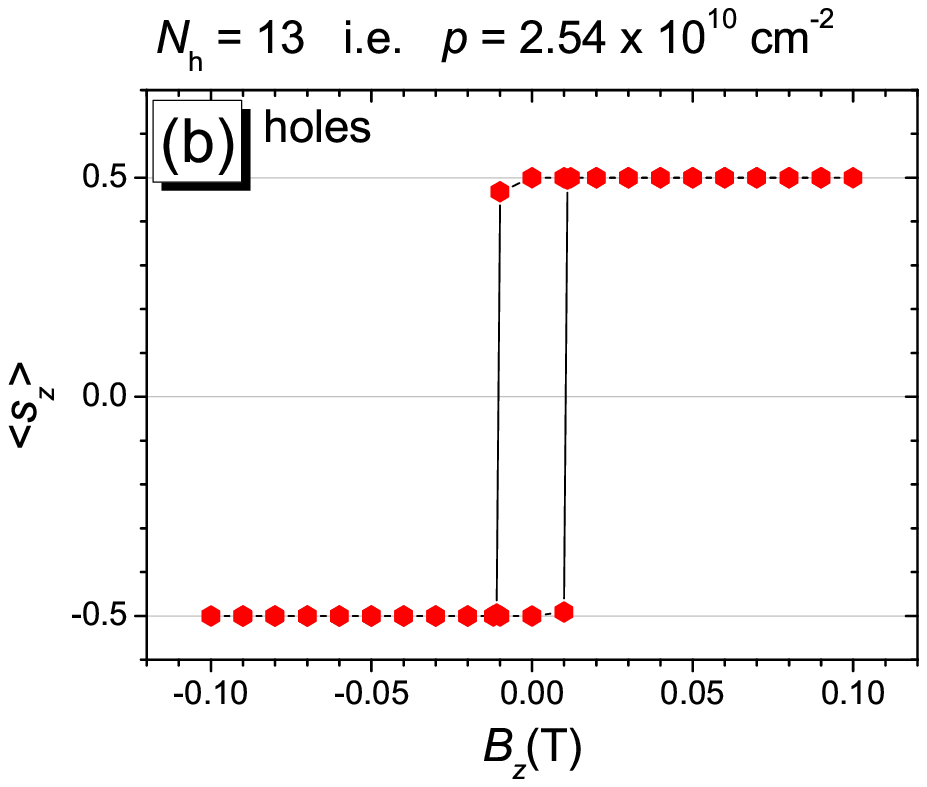}
\caption{Hysteresis loops of Mn ions and holes obtained by Monte Carlo simulations at 0.2 K
for a $p$-type quantum well (QW) containing a Cd$_{0.96}$Mn$_{0.04}$Te layer in a narrow region in the center of the QW.
The number of holes is $N_h = 13$.}
\label{fig:Nh=13_T=0.2K_delta_HL}
\end{center}
\end{figure}

\begin{figure}[t!]
\begin{center}
\includegraphics[scale=0.6]{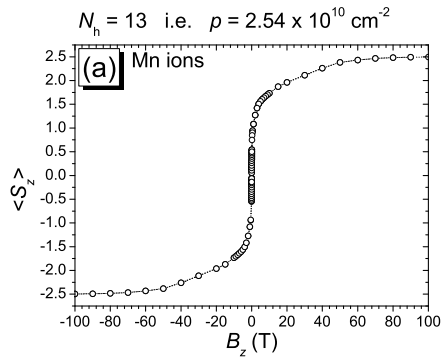}
\vspace{-0.5cm}
\caption{Hysteresis loop of Mn ions in a wide magnetic field range obtained by Monte Carlo simulations at 0.2 K for a $p$-type quantum well containing a Cd$_{0.96}$Mn$_{0.04}$Te layer in a narrow region in the center of the QW. The number of holes is $N_h = 13$.}
\label{fig:Nh=13_T=0.2K_delta_HL_large_scale}
\end{center}
\end{figure}

Here we report on simulations of hysteresis loops assuming that the Mn layer is narrower,
namely that the Mn ions are randomly distributed in a $L_{\mathrm{W}}/4$-wide region in the center of the QW.
The magnetic field is applied along the growth $z$-axis.
Our results show that narrowing of the region where Mn reside leads to an increase in remanence.
In Table~\ref{table:non-delta_delta} the values of remanent spin of Mn ions and holes
($| \langle S_z \rangle |$ and $| \langle s_z \rangle |$, respectively, for $B = $ 0), are compared
for the case of uniformly Mn-doped QWs and centrally delta-Mn-doped QWs.
In the former case Mn ions are randomly distributed all over the QW.
In the latter case Mn ions are randomly distributed in a $L_{\mathrm{W}}/4$-wide region in the center of the QW.
As seen, the remanent spin of the Mn ions is
more than two (three) times greater in the case of delta-Mn-doped QWs for Mn content $x = 4 \%$ ($8 \%$).
We see also that the magnitude of remanence at 0.2 K
and for $x \geq 4$\% diminishes with the Mn content in
the Cd$_{1-x}$Mn$_x$Te layer,
as the number of AFM pairs increases over linearly with $x$.
Furthermore, for a relatively small number of holes $N_\mathrm{h} =$ 13, i.e., $p = 2.54 \times 10^{10}$ cm$^{-2}$,
hysteresis loops are narrow and holes polarize Mn spins only partly,
as shown in Figs.~\ref{fig:Nh=13_T=0.2K_delta_HL} and \ref{fig:Nh=13_T=0.2K_delta_HL_large_scale}
presenting magnetization loops in a narrow and wide magnetic field range, respectively.

\vspace{-0.5cm}

\begin{figure}[b!]
\begin{center}
\includegraphics[scale=0.6]{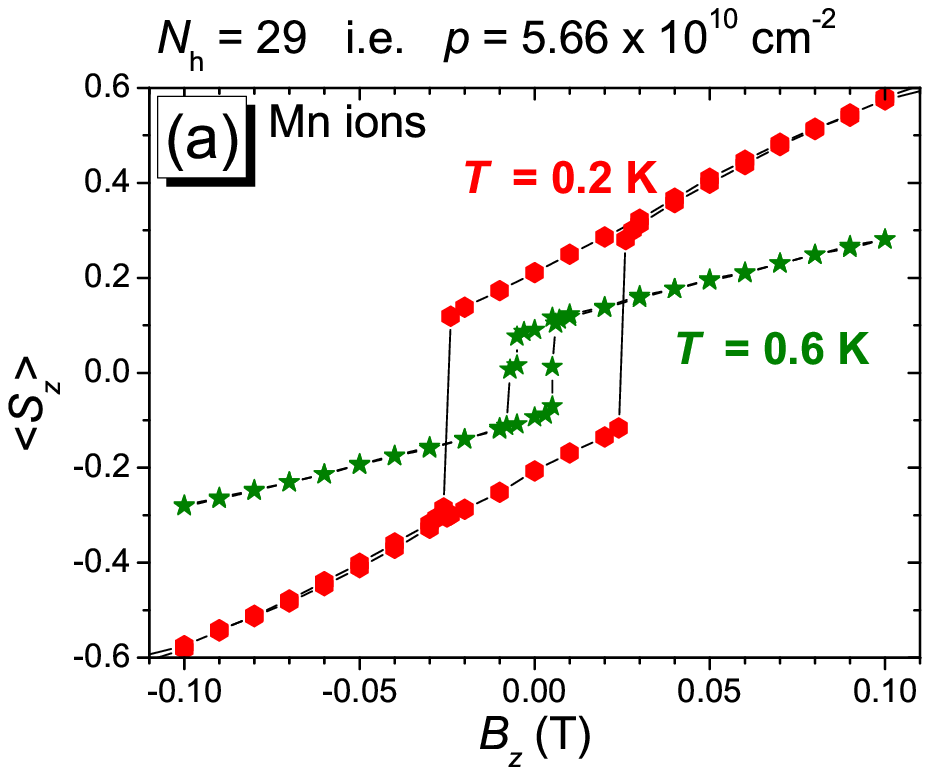}
\vspace{-0.75cm}
\includegraphics[scale=0.6]{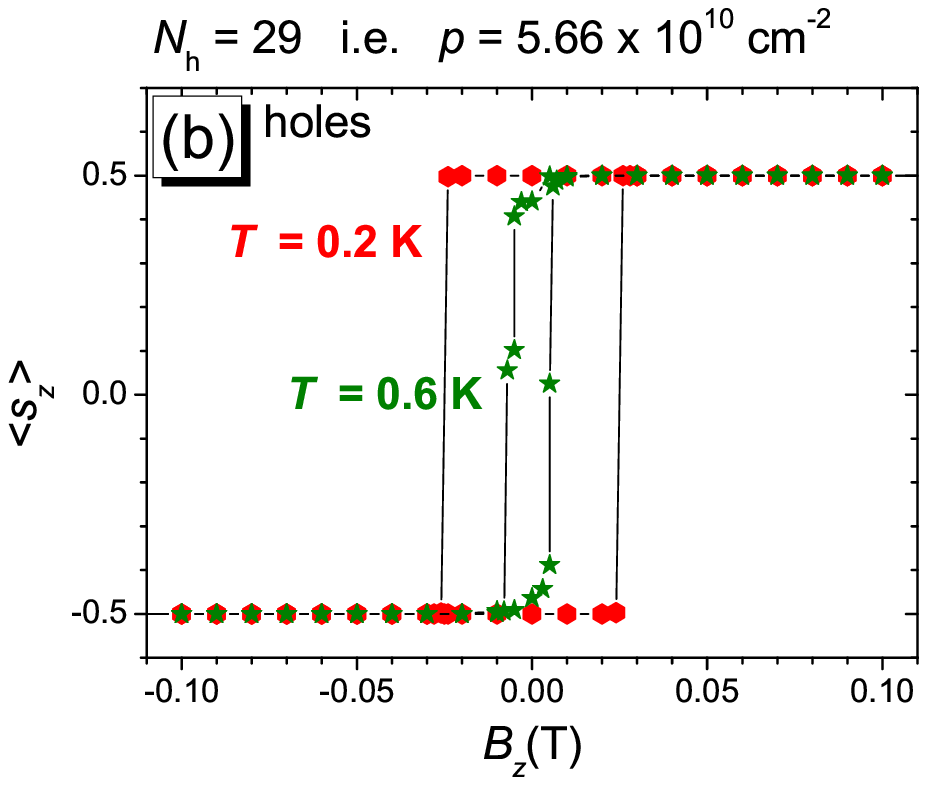}
\caption{Hysteresis loops of Mn ions and holes obtained by Monte Carlo simulations at 0.2 K and 0.6 K
for a $p$-type quantum well (QW) containing a Cd$_{0.96}$Mn$_{0.04}$Te layer in a narrow region in the center of the QW.
The number of holes is $N_h = 29$.}
\label{fig:Nh=29_T=0.2K_0.6K_delta_HL}
\end{center}
\end{figure}

We have carried out simulations with increasing number of holes.
Figures~\ref{fig:Nh=13_T=0.2K_delta_HL}-\ref{fig:Nh=29_T=0.2K_0.6K_delta_HL} present
such sizeable hysteresis loops for $N_\mathrm{h} =$ 13 and 29.
In Fig.~\ref{fig:Nh=13_T=0.2K_delta_HL},     $N_\mathrm{h} =$ 13, i.e., $p = 2.54 \times 10^{10}$ cm$^{-2}$ and $T = $ 0.2 K, while
in Fig.~\ref{fig:Nh=29_T=0.2K_0.6K_delta_HL} $N_\mathrm{h} =$ 29, i.e., $p = 5.66 \times 10^{10}$ cm$^{-2}$ and $T = $ 0.2 or 0.6 K. Further results  are summarized in Table~\ref{table:delta}.
As could be expected, we find that when $N_\mathrm{h}$ increases, the hysteresis loops become higher and wider.

\begin{table}[h!]
\caption{Remanent spin of Mn ions ($| \langle S_z \rangle |$ for $B = 0$) and coercive field (in Teslas)
obtained by Monte Carlo simulations at various temperatures and areal hole densities
for $p$-type quantum well (QW) containing a Cd$_{0.96}$Mn$_{0.04}$Te layer
in a narrow region in the center of the QW.}
\vspace{0.2cm}
\centerline{
\begin{tabular}{|c|c|c|c|c|} \hline
$T$ (K)& $N_\mathrm{h}$ & $p$ (cm$^{-2}$)         & Remanence & Coercivity \\
       &                &                         & of Mn ions&            \\ \hline \hline
0.2    & 13             & 2.54 $\times$ 10$^{10}$ & 0.11      & 0.010      \\ \hline
0.2    & 29             & 5.66 $\times$ 10$^{10}$ & 0.21      & 0.025      \\ \hline
0.6    & 29             & 5.66 $\times$ 10$^{10}$ & 0.09      & 0.005      \\ \hline
0.2    & 57             & 1.11 $\times$ 10$^{11}$ & 0.44      & 0.042      \\ \hline
\end{tabular}}
\label{table:delta}
\end{table}
\vspace{-0.5cm}

\section{Conclusion}
\label{sec:conclusion}
Our MC simulations show that by using delta-Mn-doped CdTe QWs,
where Mn ions are randomly distributed only in a $L_w/4$-wide region in the center of the QW,
enhanced remanence can be achieved
in comparison with
uniformly Mn-doped QWs where the Mn ions are randomly distributed all over the QW.
In this way sizable hysteresis loops can be achieved.
Increasing the number of holes from $N_\mathrm{h} =$ 13 to $N_\mathrm{h} =$ 57
the remanence and the coercivity is enhanced.
We presented examples of such hysteresis loops for
$N_\mathrm{h} =$ 13 at $T =$ 0.2 and
$N_\mathrm{h} =$ 29 at $T =$ 0.2 and 0.6 K.
Our results suggest that large remanence and coercivity are expected in such centrally delta-Mn-doped CdTe QWs.
Simulations are continuing in order to explore the range of material and geometrical parameters which are important to obtain hysteresis loops with desired characteristics.

\end{document}